\documentstyle[12pt]{article}

\begin{document}
{\it University of Shizuoka}

\hspace*{10cm} {\bf US-93-04}\\[-.3in]

\hspace*{10cm} {\bf February 1993}\\[.3in]

\begin{center}

{\large\bf The Rosner--Worah Type Quark Mass Matrix }\\[.1in]
{\large\bf and Ansatz of Maximal $CP$ Violation}\\[.5in]

{\bf Yoshio Koide}

Department of Physics, University of Shizuoka

52-1 Yada, Shizuoka 422, Japan \\[.1in]

and \\[.1in]

{\bf Hideo Fusaoka}

Department of Physics, Aichi Medical University

Nagakute, Aichi 480-11, Japan \\[.6in]

{\large\bf Abstract}\\[.1in]

\end{center}

\begin{quotation}
A democratic form version of the Rosner--Worah quark mass matrix is
discussed from a phenomenological point of view.
It is pointed out that an ansatz of ``maximal $CP$ violation"
can provide reasonable values of the Kobayashi-Maskawa mixings.
\end{quotation}

\vfill

To be pubulished in Phys.~Rev.~D.
\newpage
Recently, the authors [1] have determined, without taking any numerical
expression, the general form of up- and down-quark mass matrices
$(M_u, M_d)$ which are consistent with the present data on the quark masses
$m_q$ and the Kobayashi-Maskawa (KM) [2] mixing matrix $V$ as follows:
$$ M_u=U_0^\dagger D_u U_0 \ , \ \ \ M_d=U_0^\dagger \widehat{M} U_0 \ ,
\eqno(1)$$
$$ D_u \equiv
\left(
\begin{array}{ccc}
u_1 & 0 & 0 \\
0 & u_2 & 0 \\
0 & 0 & u_3
\end{array} \right) \equiv
u_3 \left(
\begin{array}{ccc}
r_{1u}\lambda^8 & 0 & 0 \\
0 & r_{2u}\lambda^4 & 0 \\
0 & 0 & 1
\end{array} \right) \ ,  \eqno(2)$$
$$ \widehat{M}\equiv V D_d V^\dagger  \simeq d_3 \left(
\begin{array}{ccc}
(r_{1d}+r_{2d}v_1^2)\lambda^4 & r_{2d} v_1 e^{i\widehat{\phi}_{12}}\lambda^3
& v_3 e^{-i\widehat{\phi}_{31}}\lambda^3  \\
r_{2d} v_1 e^{-i\widehat{\phi}_{12}}\lambda^3  & r_{2d}\lambda^2
& v_2 e^{i\widehat{\phi}_{23}}\lambda^2 \\
v_3 e^{i\widehat{\phi}_{31}}\lambda^3  & v_2e^{-i\widehat{\phi}_{23}}
\lambda^2  & 1
\end{array}\right) \ ,\eqno(3)$$
where
$$ v_1\equiv |V_{us}| /\lambda \ , \ \
 v_2\equiv |V_{cb}| /\lambda^2 \ , \ \
 v_3\equiv |V_{ub}| /\lambda^3 \ , \ \
 w\equiv (|V_{cd}|^2 - |V_{us}|^2 )/\lambda^6 \ , \eqno(4)$$
$$  r_{1d}\equiv ({d_1/d_3})/{\lambda^4} \ , \ \
  r_{2d}\equiv ({d_2/d_3})/{\lambda^2} \ ,\eqno(5)$$
$U_0$ is an arbitrary unitary matrix,
$(u_1,u_2,u_3)$ and $(d_1,d_2,d_3)$ denote quark masses $(m_u, m_c, m_t)$
and $(m_d, m_s, m_b)$, respectively, and the parameters $v_1$, $v_2$, $v_3$,
$w$, $r_{1u}$, $r_{2u}$ $r_{1d}$ and $r_{2d}$ are of the order of one.
In the expression (3), we have denoted only the first leading term for
each matrix element.
The parameter $w$ is needed
in description of the higher order terms.

The expression (3) has been derived from a general property for
$3\times 3$ Hermitian mass matrix model: in a quark basis
in which up-quark mass matrix $M_u$ takes a diagonal form $M_u=D_u$,
the seven independent parameters
in the down-quark mass matrix
$M_d=V D_d V^\dagger \equiv \widehat{M}$, i.e.,
$\widehat{M}_{11}$, $\widehat{M}_{22}$, $\widehat{M}_{33}$,
$|\widehat{M}_{12}|$, $|\widehat{M}_{23}|$, $|\widehat{M}_{31}|$, and
$$
\widehat{\phi}\equiv \widehat{\phi}_{12}+\widehat{\phi}_{23}
+\widehat{\phi}_{31} \ , \eqno(6)
$$
($\widehat{\phi}_{ij}$ are phases [3]
of the matrix elements $\widehat{M}_{ij}$)
can  completely be determined by the three down-quark masses $(d_1,d_2,d_3)$
and the four independent KM matrix parameters (4)
(it is convenient [4] to take a set of the parameters
$ \alpha\equiv |V_{us}|$,
$ \beta\equiv |V_{cb}|$, $\gamma\equiv |V_{ub}|$, and
$ \omega \equiv |V_{cd}|^2 - |V_{us}|^2$ as the four independent KM
matrix parameters).
Note that in order to determine the matrix $\widehat{M}$ we do not need
the knowledge of top quark mass.

It should be noticed that we have used only the experimental facts [5] that
$|V_{us}| \sim \lambda$, $|V_{cb}|\sim \lambda^2$,
and $|V_{ub}|\sim \lambda^3$
and $ {m_d}/{m_s}\sim {m_s}/{m_b}\sim O(\lambda^2)$, and we have not used
the explicit values of $|V_{ij}|$ and $(d_1,d_2,d_3)$, because
the current values of those are estimated model-dependently at present [6].

The matrix $\widehat{M}$, (3),
shows $\widehat{M}_{33}\gg\widehat{M}_{22}\gg\widehat{M}_{11}$,
$\widehat{M}_{22}\sim \widehat{M}_{23}$ and
$\widehat{M}_{12}\sim \widehat{M}_{31}$.
As stated below, when we transform such a mass matrix $\widehat{M}$ into
a democratic form [7], we can find some interesting relations among
quarks and KM matrix elements by putting an ansatz for violation of the
democratic form on it.

The transformation of an Hermitian matrix $\widetilde{M}$,
where $\widetilde{M}_{ij}\equiv \widetilde{m}_{ij}
e^{i\widetilde{\phi}_{ij}}$,
into a democratic type form $M$ is generally given by
the transformation matrix $A$
$$ A \equiv \left(
\begin{array}{ccc}
\frac{1}{\sqrt{2}} & -\frac{1}{\sqrt{2}} & 0 \\
\frac{1}{\sqrt{6}} & \frac{1}{\sqrt{6}} & -\frac{2}{\sqrt{6}} \\
\frac{1}{\sqrt{3}} & \frac{1}{\sqrt{3}} & \frac{1}{\sqrt{3}}
\end{array} \right) \ , \eqno(7)$$
as follows:
$$ M \equiv A^\dagger \widetilde{M} A = \widetilde{m}_{33} X
+ \widetilde{m}_{22} (-X+Y+Z) +\widetilde{m}_{11}({\bf 1}-Y-Z)$$
$$+\frac{1}{\sqrt{2}}\widetilde{m}_{23}\left[ \widetilde{c}_{23}
(-X+2Y-Z)+\widetilde{s}_{23} F(\frac{\pi}{2})
\right] $$
$$+\frac{1}{\sqrt{3}}(\widetilde{m}_{12}\widetilde{c}_{12}
+\sqrt{2} \widetilde{m}_{31} \widetilde{c}_{31}) K
+\frac{1}{\sqrt{3}}(\widetilde{m}_{12}\widetilde{s}_{12}
-\sqrt{2} \widetilde{m}_{31} \widetilde{s}_{31}) L $$
$$+\frac{1}{\sqrt{6}}\left[ \sqrt{2}\widetilde{m}_{12}
G(\widetilde{\phi}_{12})
- \widetilde{m}_{31}G(-\widetilde{\phi}_{31})\right] \ ,\eqno(8)$$
where {\bf 1} is the $3\times 3$ unit matrix,
$$
X= \frac{1}{3}\left(
\begin{array}{ccc}
1 & 1 & 1 \\
1 & 1 & 1 \\
1 & 1 & 1
\end{array} \right) \ , \ \ \
Y= \frac{1}{2}\left(
\begin{array}{ccc}
1 & 1 & 0 \\
1 & 1 & 0 \\
0 & 0 & 0
\end{array} \right) \ , \ \
Z=  \left(
\begin{array}{ccc}
0 & 0 & 0 \\
0 & 0 & 0 \\
0 & 0 & 1
\end{array} \right) \ , $$
$$
K= \left(
\begin{array}{ccc}
1 & 0 & 0 \\
0 & -1 & 0 \\
0 & 0 & 0
\end{array} \right) \ , \ \ \
L= \left(
\begin{array}{ccc}
0 & i & 0 \\
-i & 0 & 0 \\
0 & 0 & 0
\end{array} \right) \ , $$
$$
F(\delta)= \left(
\begin{array}{ccc}
0 & 0 & e^{i\delta} \\
0 & 0 & e^{i\delta} \\
e^{-i\delta} & e^{-i\delta} & 0
\end{array} \right) \ , \ \ \
G(\delta)= \left(
\begin{array}{ccc}
0 & 0 & -e^{i\delta} \\
0 & 0 & e^{i\delta} \\
-e^{-i\delta} & e^{-i\delta} & 0
\end{array} \right) \ ,  \eqno(9)$$
$\widetilde{c}_{ij}=\cos\widetilde{\phi}_{ij}$ and
$\widetilde{s}_{ij}=\sin\widetilde{\phi}_{ij}$.

We assume that when the democratic form $X$ is broken
by some additional terms, the mass matrix still reserves partially its
democratic form such as $M_{11}=M_{22}=M_{12}=M_{21}$.
The ansatz of ``partial conservation of the democratic form"
requires absence of {\bf 1}-, $K$- and $L$-terms.
The absence of {\bf 1}-term leads to the constraint
$$  \widetilde{m}_{11}=0 \ .   \eqno(10)$$
The absence of $K$- and $L$-terms leads to the constraints
$$  \widetilde{\phi}_{12}+\widetilde{\phi}_{31}=\pi \ ,  \eqno(11) $$
and
$$  \widetilde{m}_{12}=\sqrt{2}\, \widetilde{m}_{31} \ . \eqno(12) $$
Then, the general form (8) becomes a simple form
$$ M\equiv A^\dagger \widetilde{M} A = \widetilde{m}_{33} X
+ \widetilde{m}_{22} (-X+Y+Z) $$
$$+\frac{1}{\sqrt{2}}\widetilde{m}_{23}\left[ \widetilde{c}_{23}
(-X+2Y-Z)+\widetilde{s}_{23} F(\frac{\pi}{2})
\right] +\frac{\sqrt{3}}{2}\widetilde{m}_{12}G(\widetilde{\phi}_{12})
\ .\eqno(13)$$

First, in order to give a rough sketch of the model,
we apply the ansatz ((10)--(12))
to the mass matrix $A^\dagger\widehat{M} A$.
The constraint (10) gives the well-known prediction [8]
$r_{1d}+r_{2d}v_1^2\simeq 0$, i.e.,
$$   |V_{us}|\simeq \sqrt{-d_1/d_2} \simeq 0.22 \ .  \eqno(14) $$
The constraint (12) provides the relation
$v_1 r_{2d}\simeq \sqrt{2}\, v_3$ [9], i.e.,
$$
|V_{ub}| \simeq
\frac{1}{\sqrt{2}}|V_{us}|\, \frac{d_2}{d_3} \simeq 0.0044 \ . \eqno(15)$$

For the $\lambda^2$-terms ($\widetilde{m}_{12}$- and
$\widetilde{m}_{31}$-terms),
the simplest assumption which accords with the ansatz of ``partial
conservation of the democratic form"
will be Tanimoto's ansatz [10]:
the democratic type mass matrix is broken by
a ``partially" democratic type matrix $Y$ (in other words,
we assume that the absence of $Z$- and $F(\pi/2)$-terms).
The ansatz means
$$  \widetilde{m}_{22}=\widetilde{m}_{23}/\sqrt{2} \ , \eqno(16) $$
with
$$  \widetilde{\phi}_{23}=0 \ . \eqno(17) $$
For $A^\dagger \widehat{M} A$, the constraint (16) leads to the prediction
$r_{2d}\simeq v_2/\sqrt{2}$, i.e.,
$$
 |V_{cb}| \simeq \sqrt{2}d_2/d_3
\simeq 0.040 \ .  \eqno(18)$$
By combining the results (15) and (18), we can obtain a successful
relation
$$
\left|\frac{V_{ub}}{V_{cb}}\right| \simeq \frac{1}{2}|V_{us}|
\simeq 0.11 \ . \eqno(19)$$

The successful predictions (14), (15), (18) and (19) suggest
that the following quark mass
matrix form is favorable to the observed data:
$$M_q=
\frac{1}{3} a_q \left(
\begin{array}{ccc}
1 & 1 & 1 \\
1 & 1 & 1 \\
1 & 1 & 1
\end{array} \right) +
\frac{1}{2} b_q \left(
\begin{array}{ccc}
1 & 1 & 0 \\
1 & 1 & 0 \\
0 & 0 & 0
\end{array} \right) +
c_q \left(
\begin{array}{ccc}
0 & 0 & -e^{i\delta_q} \\
0 & 0 & e^{i\delta_q} \\
-e^{-i\delta_q} & e^{-i\delta_q} & 0
\end{array} \right) \ . \eqno(20)
$$
The form (20) is equivalent to the form
$$ \widetilde{M}_q\equiv A M_q A^\dagger = a_q \left(
\begin{array}{ccc}
0 & 0 & 0 \\
0 & 0 & 0 \\
0 & 0 & 1
\end{array} \right) + \frac{1}{3} b_q \left(
\begin{array}{ccc}
0 & 0 & 0 \\
0 & 1 & \sqrt{2} \\
0 & \sqrt{2} & 2
\end{array} \right)$$
$$ + \sqrt{\frac{2}{3}} c_q \left(
\begin{array}{ccc}
0 & \sqrt{2}e^{i\delta_q} & -e^{i\delta_q} \\
\sqrt{2}e^{-i\delta_q} & 0 & 0 \\
-e^{-i\delta_q} & 0 & 0
\end{array} \right) \ , \eqno(21)
$$
which  has been proposed by Rosner and Worah [11]
from a composite model of quarks.
The democratic form (20) was first
proposed by Koide [7] from the phenomenological study of the mass matrix
$A^\dagger \widehat{M} A$,
but he did not mention anything about the phase parameters $\delta_q$.
Matumoto [12] has pointed out that in the democratic form (20) the choice
$\delta\equiv\delta_u-\delta_d=\pi/2$ is the most favorable to
the experimental data.
As pointed out by Rosner and Worah and by Matumoto, indeed,
only when $\delta\equiv \delta_u-\delta_d \simeq \pi/2$, we can get
the satisfactory predictions
$$ |V_{us}|\simeq \sqrt{\frac{-d_1}{d_2}+\frac{-u_1}{u_2}} \ , \eqno(22)$$
$$ |V_{cb}|\simeq \sqrt{2}\left( \frac{d_2}{d_3}-\frac{u_2}{u_3}\right) \ ,
\eqno(23)$$
$$ \frac{|V_{ub}|}{|V_{cb}|}\simeq \frac{1}{2}|V_{us}|\simeq 0.11 \ .
\eqno(24)$$

Although Rosner and Worah, and Matumoto found that the choice
$\delta=\pi/2$ is favorable to the data, they did not give any plausible
explanation of the reason why $\delta$ takes $\pi/2$.
They have chosen the value of $\delta$ by hand such that the prediction of
$|V_{us}|$ shows a reasonable value compared with the experimental value [5]
$|V_{us}|_{exp}=0.2205\pm 0.0018$.

Note that we can choose values of the phase parameters $\delta_u$ and
$\delta_d$ independently of the quark masses $u_i$ and $d_i$, respectively.
The value of $\delta\equiv \delta_u - \delta_d$ affects only
the predictions of
$V_{ij}$, so that it does the predictions of the rephasing-invariant
quantity $J$ [13], which is a measure of $CP$ nonconservation.

Now we put an interesting ansatz which leads to the choice $\delta\simeq
\pi/2$.  In the model (20) (or (21))
the rephasing-invariant quantity $J$
is exactly given by the form
$$ J=C_0 \sin\delta ( 1+ C_1 \cos\delta) \ , \eqno(25)$$
where $C_0$ and $C_1$ are expressed approximately by
$$ C_0 \simeq 3 \sqrt{\frac{-u_1}{u_2}}\sqrt{\frac{-d_1}{d_2}}
\left( \frac{d_2}{d_3}-\frac{u_2}{u_3}\right)^2  \ , \eqno(26)$$
$$ C_1 \simeq \frac{1}{3} \sqrt{\frac{-u_1}{u_2}}\sqrt{\frac{-d_1}{d_2}}
\left(\frac{u_2/u_3}{d_2/d_3}\right)  \ . \eqno(27)$$
We require that the phase parameter $\delta\equiv\delta_u-\delta_d$ takes
such  a value $\delta_m$ as
$|J|$ takes maximal value at $\delta=\delta_m$.
Then, the values $\delta_m$ is given by
$$ \cos\delta_m = \frac{1}{4C_1}\left( 1-
\sqrt{1+8C_1^2}\right) \simeq -C_1 \ .  \eqno(28)$$
For the numerical values [6] $u_1/u_2\simeq -0.0039$,
$d_1/d_2\simeq -0.050$, $u_2/u_3\simeq 0.0042$
and $d_2/d_3\simeq 0.028$, we can obtain $C_0\simeq 2.3 \times 10^{-5}$
and $C_1 \simeq 0.80 \times 10^{-3}$ from the direct evaluation of $J$
without using the approximate reltations (26) and (27).
The value of $C_1$ leads to $\delta_m \simeq (90+0.049)^{\circ}$.
Thus, the ansatz of the maximal $CP$ violation
leads reasonably to the choice of $\delta\simeq \pi/2$.

However, it should also be noted that the case of $\delta =\pi/2$
does not mean the case of $\widehat{\phi}=\pi/2$ which corresponds to
the case of model-independent  ``maximal $CP$ violation" where
$|J|_{max}$ is given by [4]
$$|J|_{max}=\alpha\beta\gamma\sqrt{1-\frac{\alpha^2}{1-\gamma^2}}
\sqrt{1-\frac{\beta^2}{1-\gamma^2}} \ ,\eqno(29)$$
where $\alpha\equiv |V_{us}|$, $\beta\equiv |V_{cb}|$ and $\gamma\equiv
|V_{ub}|$, which are fixed at the observed values.
Since the rephasing-invariant quantity $J$ in the expression (3)
is given by
$$ J=\frac{|\widehat{M}_{12}| |\widehat{M}_{23}|
|\widehat{M}_{31}|}{(d_3-d_1)(d_3-d_2)(d_2-d_1)}
\sin\widehat{\phi} \simeq \lambda^6 v_1v_2v_3 \sin\widehat{\phi}
 \ , \eqno(30)$$
the comparison with (25) leads to the relation
$$ \sin\widehat{\phi} \simeq 3\sqrt{\frac{-u_1/u_2}{-d_1/d_2}}\sin\delta
\simeq 0.836 \times \sin\delta \ . \eqno(31)$$
Note that in the limit of $u_1/u_2=0$ the rephasing-invariant quantity $J$
becomes vanishing.

Also note that in this model the sizable value of $\sin\widehat{\phi}$
comes out
only when the up-quark mass matrix $M_u$ is given by the form (20)
as well as $M_d$.
If we take $\widehat{M}_u=D_u$ (so that $M_u\equiv A^\dagger
\widehat{M}_u A$ cannot become the type (20)),
then, from (11) and (17) we obtain
$\widehat{\phi}_{12}+\widehat{\phi}_{31}=\pi$ and $\widehat{\phi}_{23}=0$,
respectively, so that
we would meet with a wrong result $\sin\widehat{\phi}=0$.
It is essential that the ansatz is applied not to $(A^\dagger D_u A,
A^\dagger \widehat{M} A)$ in which $D_u$ does not take a same structure
with $\widehat{M}$, but to $(A^\dagger \widetilde{M_u} A,
A^\dagger \widetilde{M_d} A)$ in which $\widetilde{M}_u\equiv
R^\dagger D_u R$
takes a same structure with $\widetilde{M}_d\equiv R^\dagger \widehat{M} R$
by choosing a unitary matrix $R$ suitably.

Therefore, by introducing a small rotation
$R\equiv R(\theta_1, \theta_2, \theta_3; \chi)$
with $\theta_1=t_1\lambda^4$,
$\theta_2=t_2\lambda^6$ and $\theta_3=t_3\lambda^2$,
where $R(\theta_{1}, \theta_{2}, \theta_{3})$ is defined by
$$ R(\theta_{1}, \theta_{2}, \theta_{3}; \chi)= \left(
\begin{array}{ccc}
c_3c_2 & s_3c_2 & s_2 \\
-s_3c_1e^{i\chi}-c_3s_1s_2 & c_3c_1e^{i\chi}-s_3s_1s_2 &
s_1c_2 \\
s_3s_1e^{i\chi}-c_3c_1s_2 & -c_3s_1e^{i\chi}-s_3c_1s_2 &
c_1c_2 \\
\end{array} \right) \ ,  \eqno(32)$$
($c_{i}=\cos\theta_{i}$ and $s_{i}=\sin\theta_{i}$),
we transform $(D_u,\widehat{M})$ into $(\widetilde{M}_u,\widetilde{M}_d)$:
$$
   \widetilde{M}_u \equiv R^\dagger D_u R
= u_3 \left(
\begin{array}{ccc}
(r_{1u}+r_{2u}t_3^2)\lambda^8  &  -r_{2u} t_3 \lambda^6
& (-t_2+t_3t_1e^{-i\chi})\lambda^6 \\
-r_{2u} t_3 \lambda^6  & r_{2u}\lambda^4 & -t_1e^{-i\chi} \lambda^4 \\
(-t_2+t_3t_1e^{i\chi})\lambda^6 & -t_1e^{i\chi} \lambda^4 & 1
\end{array} \right) \ , \eqno(33) $$
$$
   \widetilde{M}_d \equiv R^\dagger \widehat{M} R
= d_3 \left(
\begin{array}{ccc}
(r_{1d}+r_{2d}v_1^2)\lambda^4  &   r_{2d}v_1e^{i\chi_{12}} \lambda^3
& v_3e^{-i\chi_{31}} \lambda^3 \\
r_{2d}v_1e^{-i\chi_{12}} \lambda^3  & r_{2d}\lambda^2
& v_2e^{i\chi_{23}} \lambda^2 \\
v_3e^{i\chi_{31}} \lambda^3 & v_2e^{-i\chi_{23}} \lambda^2 & 1
\end{array} \right) \ ,  \eqno(34)$$
where $\chi_{12}=\widehat{\phi}_{12}+\chi$, $\chi_{31}=\widehat{\phi}_{31}$,
and $\chi_{23}=\widehat{\phi}_{23}-\chi$.
Here, we have denoted only the first leading term for each matrix element.
In the matrix elements $(\widetilde{M}_d)_{11}$, $(\widetilde{M}_d)_{12}$ and
$(\widetilde{M}_d)_{31}$, the next leading terms are decreased
only by a factor $\lambda$ (and not $\lambda^2$) differently from
other matrix elements.
The more detailed expressions of $(\widetilde{M}_d)_{11}$,
$(\widetilde{M}_d)_{12}$ and $(\widetilde{M}_d)_{31}$ (in unit of $d_3$)
are as follows:
$$
  (\widetilde{M}_d)_{11}=(r_{1d}+r_{2d}v_1^2)\lambda^4 - 2\cos\chi_{12}
r_{2d}v_1t_3\lambda^5 +\cdots \ ,  \eqno(35)$$
$$
  (\widetilde{M}_d)_{12}\equiv \widetilde{m}^d_{12}
e^{i\widetilde{\phi}^d_{12}}
=r_{2d}v_1 \lambda^3 \left( e^{i\chi_{12}}
-\frac{t_3\lambda}{v_1}\right) +\cdots \ ,  \eqno(36)$$
$$
  (\widetilde{M}_d)_{31}\equiv \widetilde{m}^d_{31}
e^{i\widetilde{\phi}^d_{31}}
=v_3 \lambda^3 \left( e^{i\chi_{31}}
- \frac{v_2t_3\lambda}{v_3}e^{-i\chi_{23}}\right) +\cdots \ .\eqno(37)$$
Note that $\widetilde{\phi}^d_{12}+\widetilde{\phi}^d_{31}
+\widetilde{\phi}^d_{23} \neq \widehat{\phi}_{12}
+\widehat{\phi}_{31}+\widehat{\phi}_{23}$, although
 $\chi_{12}+\chi_{31}+\chi_{23} = \widehat{\phi}_{12}
+\widehat{\phi}_{31}+\widehat{\phi}_{23}$.

The requirement $(\widetilde{M}_u)_{11}=0$ leads to
$   t_3 \simeq \pm \sqrt{-r_{1u}/r_{2u}}$,
and the requirement $(\widetilde{M}_d)_{11}=0$ leads to
$$
  |V_{us}|\simeq \sqrt{-\frac{d_1}{d_2}} \pm\cos\chi_{12}
\sqrt{-\frac{u_1}{u_2}} \ .  \eqno(38)$$
By putting (36) and (37) into the constraints
$\widetilde{m}^d_{12}=\sqrt{2}\,\widetilde{m}^d_{31}$
and $\widetilde{\phi}^d_{12}+\widetilde{\phi}^d_{31}=\pi$,
we can again obtain the relation (31),
where we have used $\cos^2\chi_{12}\ll 1$ which is
phenomenologically required from (35).

In conclusion, by suggested from the quark mass matrix form (3) which
was determined phenomenologically, we have reached a democratic type
quark mass matrix form (20), which is equivalent to the Rosner--Worah
quark mass matrix (21).
We have pointed out that the ansatz of ``maximal $CP$-violation" can
lead to the choice of $\delta\simeq \pi/2$, which is needed in order to
provide reasonable predictions of $|V_{ij}|$ in the model (20).

\vglue.3in
\begin{center}
{\bf Acknowledgments}
\end{center}

The authors would like to thank Professor K.~Matumoto
for calling his attention to the references [11] and [12].
This work has financially been supported in part by a Shizuoka Prefectural
Government Grant.

\vglue.4in
\newcounter{0000}
\centerline{\bf References and Footnotes}
\begin{list}
{[~\arabic{0000}~]}{\usecounter{0000}
\labelwidth=1cm\labelsep=.3cm\setlength{\leftmargin=1.7cm}
{\rightmargin=.4cm}}

\item Y.~Koide, H.~Fusaoka and C.~Habe, Phys.~Rev. {\bf D46}, R4813 (1992).
\item M.~Kobayashi and T.~Maskawa, Prog.~Theor.~Phys. {\bf 49}, 652 (1973).
\item
Note that each phase factor $\widehat{\phi}_{ij}$ is not observable quantity,
but only the phase parameter $\widehat{\phi}$ is observable, e.g.,
$CP$ nonconservation effects appear only through
the phase parameter $\widehat{\phi}$.
\item For instance, Y.~Koide, Mod.~Phys.~Lett. {\bf 6}, 2793 (1991);
{\bf 7}, 1691 (1992);
G.~B\'{e}langer, C.~Hamzaoui and Y.~Koide, Phys.~Rev. {\bf D45}, 4186 (1992).
\item Particle Data Group, K.~Hikasa {\it et al}., Phys.~Rev. {\bf D45},
S1 (1992).
\item At present, the quark mass values at an energy scale 1 GeV are not
so precisely established.  Some attempts to re-estimate quark masses
have recently been reported: for example, C.~A.~Dominguez and N.~Paver,
Phys.~Lett. {\bf B293}, 197 (1992);
J.~F.~Donoghue and B.~E.~Holstein, Phys.~Rev.~Lett. {\bf 69}, 3444 (1992).
Since  the numerical values are not so rigidly
essential in the present paper,  we, for convenience, use values
given in Ref.~[1]
as quark mass values at an energy scale 1 GeV.
\item H.~Harari, H.~Haut and J.~Weyers, Phys.~Lett. {\bf B78},
459 (1978);
Y.~Koide, Phys.~Rev.~Lett. {\bf 47}, 1241 (1981); Phys.~Rev.
{\bf D28}, 252 (1983); {\bf 39}, 1391 (1989);
C.~Jarlskog, in {\it Proceedings of the International Symposium on
Production and Decays of Heavy Hadrons}, Heidelberg, Germany, 1986
edited by K.~R.~Schubert and R. Waldi (DESY, Hamburg), p.331, 1986;
P.~Kaus, S.~Meshkov, Mod.~Phys.~Lett. {\bf A3}, 1251 (1988);
L.~Lavoura, Phys.~Lett. {\bf B228}, 245 (1989);
M.~Tanimoto, Phys.~Rev. {\bf D41}, 1586 (1990);
H.~Fritzsch and J.~Plankl, Phys.~Lett. {\bf B237}, 451 (1990).
\item S.~Weinberg, Ann.~N.~Y.~Acad.~Sci. {\bf 38}, 185 (1977);
H.~Fritzsch, Phys.~Lett. {\bf B73}, 317 (1978); Nucl.~Phys.
{\bf B115}, 189 (1979);
H.~Georgi and D.~V.~Nanopoulos, {\it ibid}. {\bf B185}, 52 (1979).
\item Y.~Koide, in {\it International Symposium on Extended Objects and
Bound Systems}, Proceedings, Karuizawa, Japan, 1992 (World Scientific,
Singapore, in press).
\item M.~Tanimoto, Phys.~Rev. {\bf D41}, 1586 (1990).
\item J.~L.~Rosner and M.~Worah, Phys.~Rev. {\bf D46}, 1131 (1992).
\item K.~Matumoto, Preprint No.~TOYAMA-75, to be published in
Prog.~Theor. Phys. {\bf 89}, No.~1, (1993).
\item C.~Jarlskog, Phys.~Rev.~Lett. {\bf 55}, 1839 (1985);
O.~W.~Greenberg, Phys.~Rev. {\bf D32}, 1841 (1985);
I.~Dunietz, O.~W.~Greenberg, and D.-d.~Wu, Phys.~Rev.~Lett. {\bf 55},
2935 (1985);
C.~Hamzaoui and A.~Barroso, Phys.~Lett. {\bf 154B}, 202 (1985);
D.-d.~Wu, Phys.~Rev. {\bf D33}, 860 (1986).

\end{list}

\end{document}